\newcommand{\BaCdV} {BaCdVO(PO$_4$)$_2$\xspace}
\newcommand{\PbV} {Pb$_{2}$VO(PO$_4$)$_2$\xspace}

\documentclass[aps,prl,reprint,showpacs,longbibliography,superscriptaddress, citeonline]{revtex4-1}
\usepackage{amsmath,amssymb,bm}
\usepackage{graphicx}
\usepackage{xspace}
\usepackage{latexsym}
\usepackage{color}
\usepackage{hyperref}
\usepackage{placeins}

\begin{document}
	\title{Presaturation  phase with no dipolar order in a  quantum ferro-antiferromagnet}
	
	\author{V.~K.~Bhartiya}
	\email{vvivek@phys.ethz.ch}
	\affiliation{Laboratory for Solid State Physics, ETH Z\"{u}rich, 8093 Z\"{u}rich, Switzerland}

	\author{K.~Yu.~Povarov}
	\affiliation{Laboratory for Solid State Physics, ETH Z\"{u}rich, 8093 Z\"{u}rich, Switzerland}

	\author{D.~Blosser}
	\affiliation{Laboratory for Solid State Physics, ETH Z\"{u}rich, 8093 Z\"{u}rich, Switzerland}
	
	\author{S.~Bettler}
	\affiliation{Laboratory for Solid State Physics, ETH Z\"{u}rich, 8093 Z\"{u}rich, Switzerland}
	
	\author{Z.~Yan}
	\affiliation{Laboratory for Solid State Physics, ETH Z\"{u}rich, 8093 Z\"{u}rich, Switzerland}
	
	\author{S.~Gvasaliya}
	\affiliation{Laboratory for Solid State Physics, ETH Z\"{u}rich, 8093 Z\"{u}rich, Switzerland}
	
	\author{S.~Raymond}
	\affiliation{Univ. Grenoble Alpes, CEA, IRIG/DEPHY/MEM-MDN, F-38000 Grenoble, France}
	
	\author{E.~Ressouche}
	\affiliation{Univ. Grenoble Alpes, CEA, IRIG/DEPHY/MEM-MDN, F-38000 Grenoble, France}
	
	\author{K.~Beauvois}
	\affiliation{Institut Laue Langevin, 38000 Grenoble, France}
	
	\author{J.~Xu}
	\affiliation{Helmholtz-Zentrum Berlin f\"{u}r Materialien und Energie GmbH, Hahn-Meitner-Platz 1, D-14109 Berlin, Germany}
	
	\author{F.~Yokaichiya}
	\affiliation{Helmholtz-Zentrum Berlin f\"{u}r Materialien und Energie GmbH, Hahn-Meitner-Platz 1, D-14109 Berlin, Germany}
	
	\author{A.~Zheludev}
	\email{zhelud@ethz.ch}
	\homepage{http://www.neutron.ethz.ch/}
	\affiliation{Laboratory for Solid State Physics, ETH Z\"{u}rich, 8093 Z\"{u}rich, Switzerland}

	\begin{abstract}
		Magnetization, magnetocaloric, calorimetric, neutron and X-ray diffraction and inelastic neutron scattering measurements are performed on single crystals of \BaCdV. The low-temperature crystal structure is found to be of a lower symmetry than previously assumed. The result is a more complicated model spin Hamiltonian, which we infer from measurements of the spin wave dispersion spectrum. The main finding is a novel spin state which emerges in high magnetic fields after antiferromagnetic order is terminated at $H_{c1}\simeq 4.0$~T.
		It is a distinct thermodynamic phase with a well-defined phase boundary at $H_{c2}\simeq 6.5$~T and is clearly separate from the fully saturated phase.
		Yet, it shows no conventional (dipolar)  magnetic long range order. We argue that it is fully  consistent with the expectations for a quantum bond-nematic state.
	\end{abstract}
	
	\date{\today}
	\maketitle
	
	Conventional long range order in spin systems is represented by a
	static, usually  periodically modulated magnetization. It breaks
	both rotational and time-reversal symmetries of the underlying
	Heisenberg Hamiltonian. Spin nematic order, as first envisioned by M.~Blume~\cite{BlumeHsieh_JApplPhys_1969_SpinNematic} and generalized by
	A.~F.~Andreev~\cite{AndreevGrishchuk_JETP_1984_SpinNematics}, breaks
	rotational but not time reversal symmetries. Spins continue to
	fluctuate, but these fluctuations \emph{spontaneously} become
	anisotropic. From a theorist's perspective such exotic quantum
	states are robust and not particularly rare, often realized in partially magnetized systems with competing antiferromagnetic (AF) and ferromagnetic (FM) interactions~\cite{Sudan_PRB_2009_ChainMultipoles,ZhitomirskyTsunetsugu_EPL_2010_nematic,BalentsStarykh_PRL_2016_QuantumLifshitz}. For example, the simple
	next-nearest-neighbor (nnn) Heisenberg AF on a
	square lattice with nearest-neighbor (nn) FM
	coupling is predicted to always be a spin nematic in some range
	of applied magnetic fields just below full polarization (FP)~\cite{ShannonMomoi_PRL_2006_J1J2squarecircle,Ueda_JPSJ_2015_NematicInField}.
	Finding an experimental realization of  even this
	relatively unconstrained model is a formidable challenge. The only known
	potential host compounds are layered vanadophosphates of type
	$AA'$VO(PO$_4$)$_2$~\cite{TsirlinRosner_PRB_2009_FSQLvanadatesSummary,Tsirlin_PRB_2009_FSQLmagnetization}.
	Until recently, no unambiguous signatures of spin nematic phases  have been
	found in any of them. The main obstacle has been a lack of single
	crystal samples. Indeed, powder data are notoriously difficult to
	interpret. In applied magnetic fields, where each
	crystallite experiences \emph{a priori} different conditions, phase
	transitions and other anomalies that may be indicative of the
	nematic state become washed out. Thus, single crystal experiments are key.

	In our recent
	paper~\cite{PovarovBhartiya_PRB_2019_BaCdVthermodynamics} we
	reported the first thermodynamic evidence of a novel presaturation quantum phase in small single crystals of \BaCdV, one of the most promising host
	materials~\cite{Smerald_PRB_2015_INSnematic}. Powder magnetometric and neutron diffraction studies by other authors
	\cite{Skoulatos_PRB_2019_PutativeNematic} seemingly supported the notion of the new phase
	being the elusive spin nematic state, but, as argued below, may otherwise be misleading. In the present work we settle this issue decisively. We report combined neutron and X-ray
	diffraction, inelastic neutron scattering, Faraday balance magnetometry, calorimetry and
	magnetocaloric(MCE) studies on appropriate size  \BaCdV\ \emph{single crystals}.  We first show that the low-temperature crystallographic structure
	is of a lower symmetry than previously assumed. Consequently, the spin wave spectrum is quite different from that of a simple square lattice. Despite that, the putative pre-saturation nematic phase exists and persists over a surprisingly wide range of applied
	magnetic fields. It has distinct thermodynamic phase
	boundaries, but its polarization is
	almost complete, over 98\% of full saturation.

	At room temperature \BaCdV\ has an orthorhombic $P_{bca}$ ($D^{15}_{2h}$, No.~$61$)
	structure with 8 V$^{4+}$ ions per unit cell
	\cite{Meyer_ZNatur_1997_CrystVanadates} forming an approximate  $S=1/2$ square lattice in the $(a,b)$ plane.
	If what is known about the
	structurally similar \PbV is any guidance
	\cite{BettlerLandolt_PRB_2019_LedVanadateMAX}, magnetic interactions
	are in the  these planes, with negligible
	inter-layer coupling. The nn and nnn in-plane exchange constants for \BaCdV\ have been
	previously estimated from powder data as $J_{1}=-0.31$ and
	$J_{2}=0.28$~meV~\cite{NathTsirlin_PRB_2008_BaCdVO(PO4)2}. There is also a weak easy-axis anisotropy that manifests in a spin
	flop transition (SF) around $\mu_{0}H_{\text{SF}}\simeq0.5$~T for
	$\mathbf{H}\parallel\mathbf{a}$~\cite{PovarovBhartiya_PRB_2019_BaCdVthermodynamics}.

\begin{figure}
		\centering
		\includegraphics[width=0.48\textwidth]{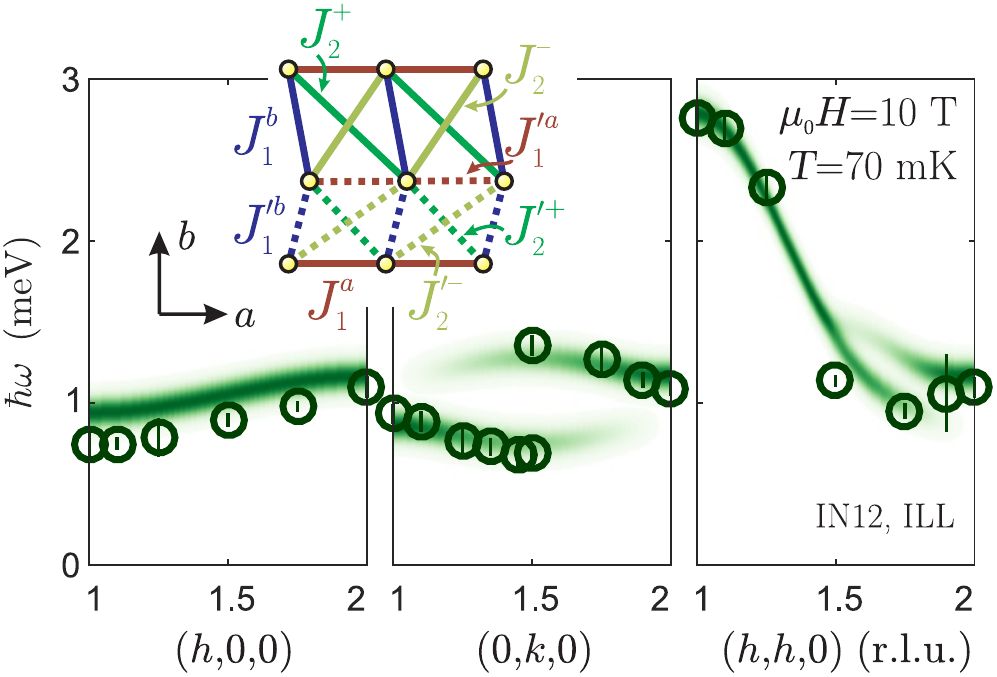}
		\caption{Spin wave spectrum in fully polarized \BaCdV. Symbols are experimental data, the shading in the background shows the  intensities from linear spin wave theory. Inset: schematic of of symmetry-allowed Heisenberg exchange interactions in the $P_{ca}2_{1}$ structural phase of \BaCdV.}\label{FIG:spinwaves}
	\end{figure}

	Long range magnetic order in zero applied field is of a peculiar ``up-up-down-down'' type, propagation vector
	$(0,1/2,0)$, and appears below $T_{N}=1.05$~K  ~\cite{Skoulatos_PRB_2019_PutativeNematic}.
	It is incompatible with the simple $J_1$-$J_2$ model and was attributed to
	additional 3rd-nearest neighbor (3nn)
	interactions in the V$^{4+}$ layers~\cite{Skoulatos_PRB_2019_PutativeNematic,Sindzingre_JPConf_2009_J1J2J3first,*Sindzingre_JPConf_2010_J1J2J3final}. There is, however, a more natural explanation. It turns out that \BaCdV undergoes a structural phase transition at $T\sim 250$~K.
	It was detected by tracking the temperature dependence of the $(0,-3,-7)$ neutron Bragg reflection with CRG-D23 diffractometer at ILL. This Bragg reflection is forbidden in  $P_{bca}$ but becomes visible below the transition point (see supplementary material).
	At low temperature the compound remains orthorhombic with the reduced space group $P_{ca}2_{1}$ ($C^{5}_{2v}$, No.~$29$).  The structure was solved at $T=120$~K from $14519$ Bragg peak intensities collected in a single crystal
	X-ray diffraction experiment using a Bruker APEX-II diffractometer.
	Details of the crystallographic refinement are reported in the supplement. The  $P_{ca}2_{1}$ structure, with two inequivalent V$^{4+}$ positions, allows for as many as 4 nn and 4 nnn interactions within each V$^{4+}$ layer as shown in the inset of Fig.~\ref{FIG:spinwaves}.
	In particular, unlike at room temperature, there is a possibility of alternating nn and nnn interactions along the crystallographic $b$ axis. This naturally explains the $(0,1/2,0)$ magnetic propagation vector without invoking 3nn coupling, which would have to span over an unrealistic $8$~\AA\ distance.

	To elucidate the effect of lowered crystal symmetry on the coupling constants in \BaCdV  we performed preliminary inelastic neutron  scattering measurements of the spin wave (SW) dispersion in the fully saturated phase. The data were collected at $T=70$~mK  on the IN12 cold neutron 3-axis  spectrometer at CRG-ILL in an $\mu_{0}H=10$~T magnetic field applied along the crystallographic $c$ axis. We co-aligned two crystals of total mass 320~mg and a combined mosaic of $10^\circ$ full width at half height grown using  98.8\% isotropically enriched 144Cd to reduce neutron absorption. A series of constant-$\mathbf{Q}$ scans performed with $3.5$~meV final energy neutrons (see supplement) reveal the dispersion relations plotted in Fig.~\ref{FIG:spinwaves} (symbols). It was analyzed using linear spin wave theory (\textsc{SpinW} package~\cite{TothLake_JPCM_2015_SpinW}). The entire set of eight exchange constants can {\it not} be uniquely determined from the three sets of measured dispersion curves. However, we find that without an alternation of
	$J$s along the $b$ direction the  two distinct spin wave branches clearly visible in the data can not be accounted for. With such an alternation, even a restricted minimal model reproduces the measured spectrum reasonable well with $J_{1}^{a}=J_{1}^{'a}=J_{1}^{b}=-0.42$, $J_{1}^{'b}=-0.34$, $J^{+}_{2}=J_{2}^{-}=0.16$ and $J_{2}^{'+}=J_{2}^{'-}=0.38$~meV. The corresponding calculated scattering intensities~\cite{TothLake_JPCM_2015_SpinW} are shown as a false color plot in  Fig.~\ref{FIG:spinwaves}. Further measurements will be needed to uniquely determine the Hamiltonian, but it is already clear that for \BaCdV the simple $J_{1}$-$J_{2}$ model is inadequate and that an alternation of exchange constants plays a crucial role.

\begin{figure}[!h]
		\includegraphics[width=0.48\textwidth]{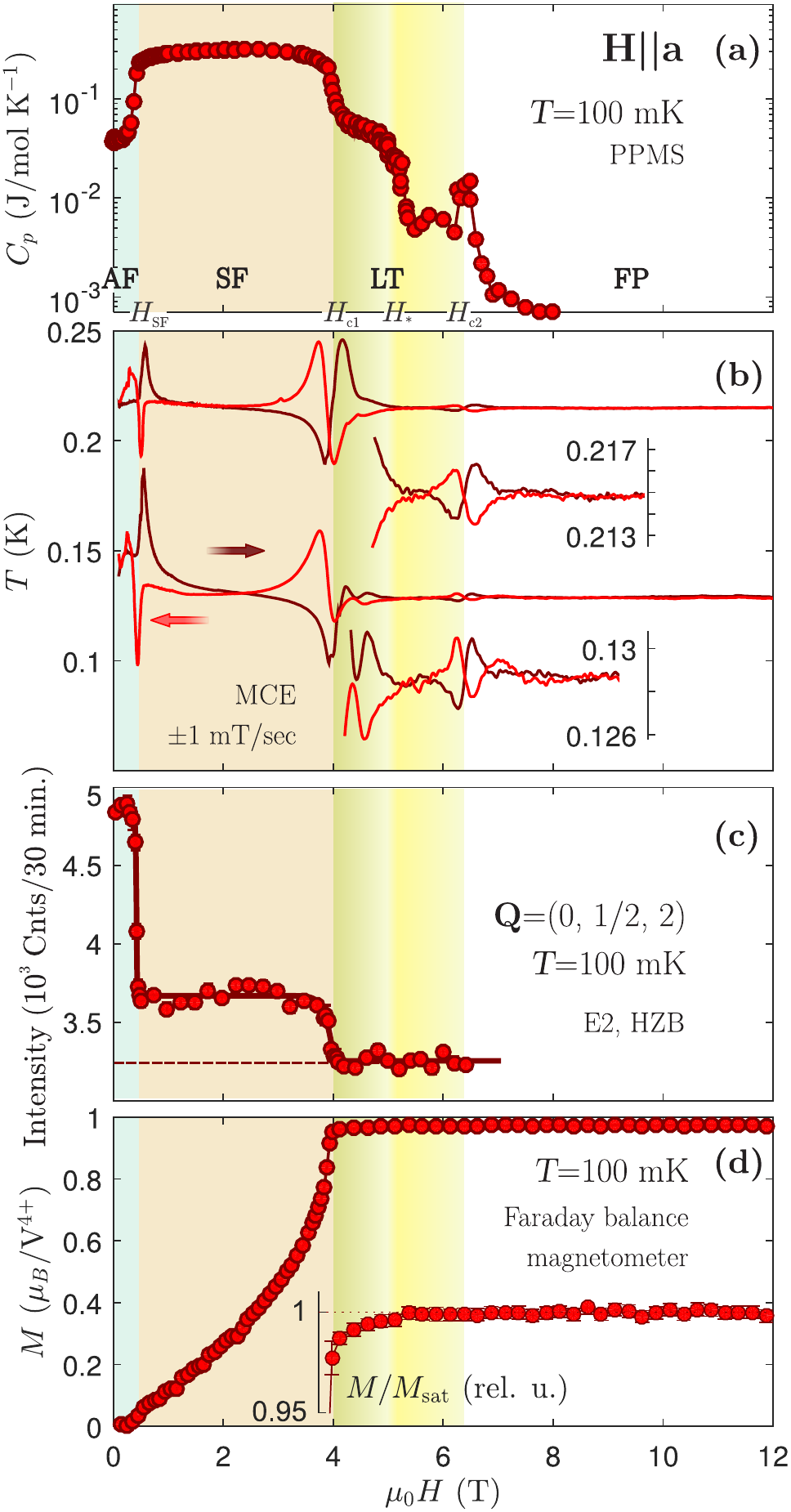}\\
		\caption{Magnetic thermodynamics of \BaCdV single crystals in the axial $\mathbf{H}\parallel\mathbf{a}$ geometry.
			(a) Specific heat at $100$~mK.
			(b) Magnetocaloric effect measurements at $130$ and $215$~mK. The insets shows the fine structure of $T(H)$ curves above
			$H_{c1}$ in more detail.
			(c) Neutron diffraction intensity at the $(0,~1/2,~2)$
			magnetic Bragg peak position at $100$~mK. The solid line is a guide to the
			eye, the dashed line shows the background intensity.
			(d) Magnetization curve at $100$~mK. The inset shows a zoomed in structure of $M/M_{sat}$ curve above
			$H_{c1}$. }\label{FIG:panelsA}
	\end{figure}

\begin{figure}[!h]
		\includegraphics[width=0.48\textwidth]{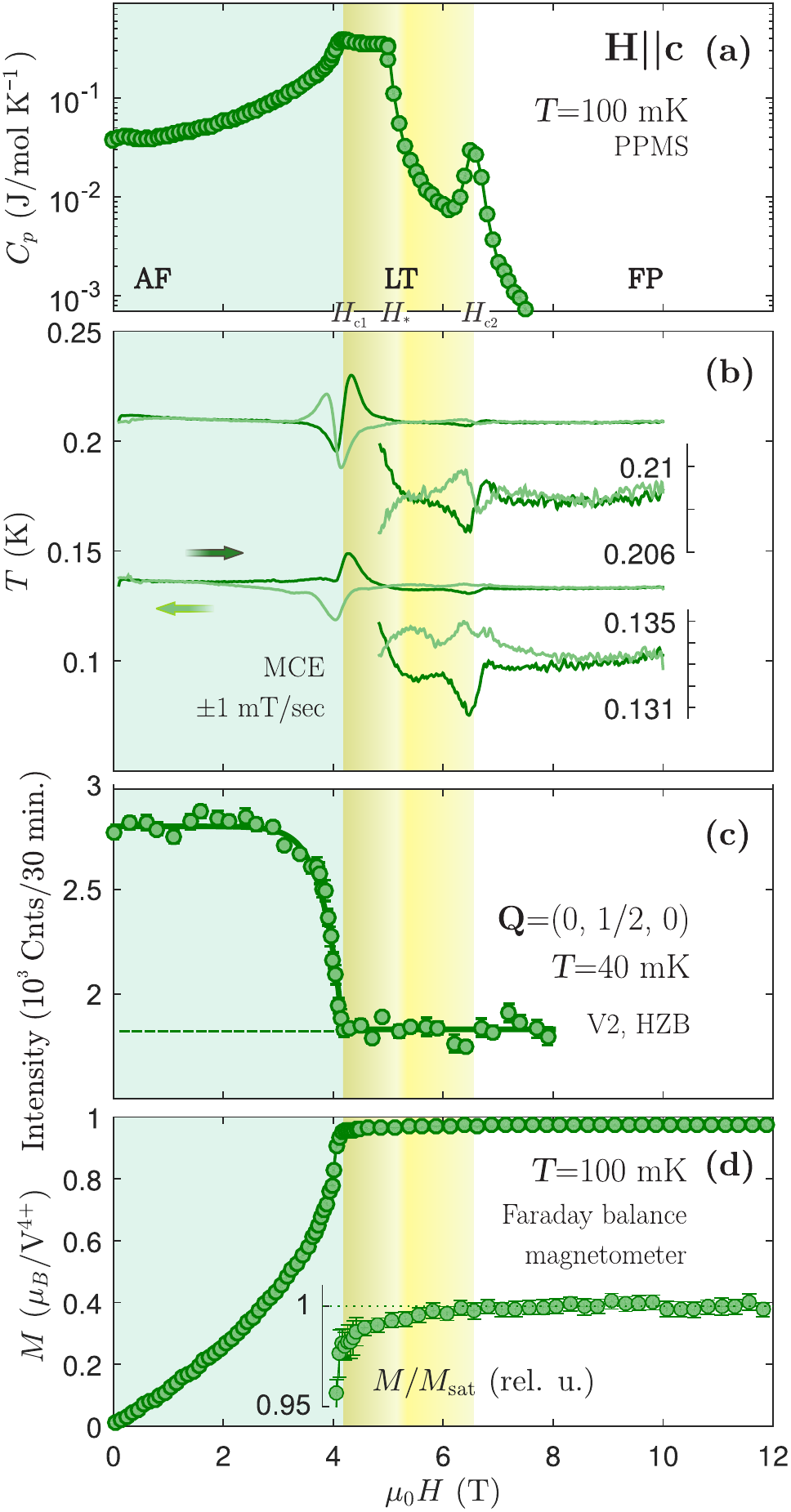}\\
		\caption{Similar measurements in the transverse $\mathbf{H}\parallel\mathbf{c}$ geometry. (c) shows the field dependence of the $(0,~1/2,~0)$ magnetic Bragg peak at 40~mK.
		}\label{FIG:panelsC}
	\end{figure}

	We now turn to the main focus of the present paper, which is the putative low-temperature  nematic phase (LT) in applied fields.
	It emerges at
	$\mu_{0}H_{c1}\simeq4$~T in either a first- or second-order transition for
	$\mathbf{H}\parallel\mathbf{a}$ and $\mathbf{H}\parallel\mathbf{b}$,
	respectively~\cite{PovarovBhartiya_PRB_2019_BaCdVthermodynamics}.
	The problem with the nematic order parameter is that it is {\em not
		directly accessible} to any experimental technique~\footnote{In
		principle, it can be probed with resonant magnetic X-ray
		diffraction. Unfortunately, this technique is incompatible with the
		high magnetic fields and mK temperatures at which the nematic state
		occurs in \BaCdV.}. The only way to identify it is thus
	by a combination of complementary measurements. Figures
	\ref{FIG:panelsA} and \ref{FIG:panelsC} show such data for two
	sample geometries, with the magnetic field applied parallel and
	perpendicular to the magnetic easy axis, respectively. All these
	measurements are carried out at base temperatures of $^3$He-$^4$He
	dilution refrigerators ($100$~mK or lower) on single crystals grown
	as described in Ref.~\cite{PovarovBhartiya_PRB_2019_BaCdVthermodynamics}.
	
	Exhibit one (top panels in Figs.~\ref{FIG:panelsA}
	and~\ref{FIG:panelsC}) is heat capacity measured vs. applied
	magnetic field using a commercial Quantum Design PPMS relaxation calorimetry setup on a $2.3$~mg
	crystal of \BaCdV. As described in detail in
	Ref.~\cite{PovarovBhartiya_PRB_2019_BaCdVthermodynamics}, for
	$\mathbf{H}\parallel\mathbf{a}$, the discontinuous spin flop transition at
	$\mu_{0}H_{\text{SF}}$ and another one at $\mu_{0}H_{c1}=4.00(5)$~T are
	marked by sharp steps in the heat capacity. For
	$\mathbf{H}\parallel\mathbf{c}$ there is no spin flop transition and
	the specific heat exhibits a strong divergence approaching
	$\mu_{0}H_{c1}=4.08(5)$~T. In both cases, fluctuations persist above
	$\mu_{0}H_{c1}$, marking the novel pre-saturation state. In all the orientations the specific heat plateau above $\mu_{0}H_{c1}$ visibly collapses around
	a crossover field $\mu_{0}H_{\ast}\simeq5.5$~T. This pronounced change in the
	resulting specific heat stems from an onset of ``temperature
	arrest'' kind of behavior in the relaxation curves. We interpret it
	as a hint of the phase separation taking place in the sample~\cite{RoesslerCherian_PRB_2011_FeTetemperaturearrest,UedaMomoi_PRB_2013_NematicSeparation}.  At any orientation at still higher field $\mu_{0}H_{c2}\simeq6.5$~T there
	clearly is an additional lambda anomaly, albeit a weak one~\footnote{In both orientations the relaxation curves for $H > H_{*}$ were taken with $\sim 30$~sec. pulse length and interpreted within a simple exponential relaxation (single $\tau$) model.}.

	The transition at $H_{c2}$ is even more obvious in exhibit
	two [panels (b) in Figs.~\ref{FIG:panelsA} and~\ref{FIG:panelsC}].
	These are magnetocaloric measurements on the same sample using a
	constant field sweep rate of $1$~mT/sec. Up and down sweeps are
	shown as darker and lighter lines, respectively. For $\mathbf{H}\parallel\mathbf{a}$ at around $130$~mK, $H_{\text{SF}}$
	and $H_{c1}$ are both marked by asymmetric
	anomalies, characteristic of discontinuous
	transitions~\cite{KohamaMarcenat_RevSciInstr_2010_MCEtheoryPulsed}.
	For  $\mathbf{H}\parallel\mathbf{c}$ the lowest-temperature magnetocaloric sweeps the anomaly at $H_{c1}$ is also asymmetric.
	At the same time, a weak but clear magnetocaloric anomaly is also present at $H_{c2}$ for all the orientations.
	At slightly elevated temperatures [$T\simeq210$~mK set of curves in Figs.~\ref{FIG:panelsA}(b) and~\ref{FIG:panelsC}(b)] the main anomaly at $H_{c1}$ turns symmetric, indicative of a continuous phase
	transition~\cite{KohamaMarcenat_RevSciInstr_2010_MCEtheoryPulsed}.
	In both geometries, at slightly elevated temperatures the transition at $H_{c2}$ is clearly
	continuous as well.
	
	\begin{figure}
		\includegraphics[width=0.48\textwidth]{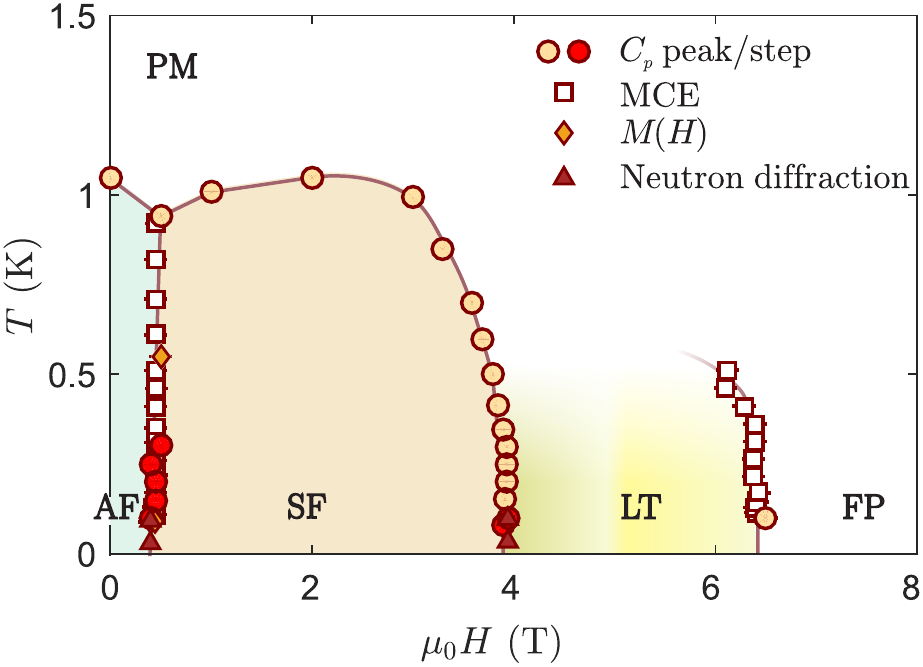}\\
		\caption{Magnetic phase diagram of \BaCdV for $\mathbf{H}\parallel\mathbf{a}$ field direction obtained by a combination of measurement techniques. Solid lines are guide to the eye. The phases are: PM and FP --- paramagnet and fully polarized paramagnet, AF and SF --- conventional antiferromagnetic states before and after spin-flop, LT --- novel presaturation phase presumed to be the quantum spin nematic state. }\label{FIG:PhD}
	\end{figure}
	
	Additional magnetocaloric effect scans at different temperatures allow us to trace the $H_{c2}$ phase boundary vs. temperature and
	reconstruct the $H-T$ phase diagram. The transition can be followed up to about
	$500$~mK. Up to $350$~mK it barely shifts and shows no sign of
	broadening: only its magnitude decreases. Above $350$~mK  the left shoulder of the magnetocaloric anomaly vanishes
	and it turns into a broad peak. This peak becomes
	indiscernible in the data noise above $500$~mK.  Fig.~\ref{FIG:PhD} shows the phase boundaries deduced from MCE, magnetization, neutron diffraction, and specific heat~\cite{PovarovBhartiya_PRB_2019_BaCdVthermodynamics}.
	Very similar temperature-dependent behavior was observed for
	$\mathbf{H}\parallel\mathbf{c}$ and $\mathbf{H}\parallel\mathbf{b}$
	(not shown). The measured phase diagram for \BaCdV shows certain similarities  to the one recently reported for volbortite, a very complicated frustrated magnet that is to some extent equivalent to a $S=1/2$ model on a frustrated square lattice and also expected to host a spin nematic phase~\cite{KohamaIshikawa_PNAS_2019_VolborthiteNematic}.

	Which of the  transitions observed in \BaCdV corresponds to a destruction of
	conventional long range order? This is clarified by exhibit 3
	[panels (c) in Figs.~\ref{FIG:panelsA} and~\ref{FIG:panelsC}], which
	shows the field dependence of the magnetic Bragg peaks $(0,1/2,2)$ and $(0,1/2,0)$ measured via
	neutron diffraction. These data were taken on the V2
	[Fig. \ref{FIG:panelsC}(c)] and E2 [Fig. \ref{FIG:panelsA}c]
	instruments at the BER reactor at HZB using $2.74$~meV and $14.2$~meV neutrons and $142$ and $24$~mg 144Cd-enriched single crystals, respectively.
	In both geometries  the disappearance of conventional magnetic order
	exactly coincides with $\mu_{0}H_{c1}$. That no magnetic structure with a different propagation vector exists in higher fields is clear from powder diffraction experiments of Ref.~\cite{Skoulatos_PRB_2019_PutativeNematic}. We thus identify the region
	$H>H_{c2}$ as fully saturated and the field range $H_{c1}<H<H_{c2}$
	as corresponding to the potential spin-nematic phase.
	
	%\begin{figure}
	%  \includegraphics[width=0.48\textwidth]{F4_Msat.pdf}\\
	%  \caption{Magnetization comparison between the axial (top) and transverse (bottom) geometries of \BaCdV in the $H_{c1}-H_{c2}$ region at
	%  $T=100$~mK. The full saturation value is estimated from the data above $8$~T. Dashed lines indicate $1.00$ and $0.98$ of $M/M_{\text{sat}}$. {\color{red}Remove the 0.98 dashed line.}}\label{FIG:DeltaM}
	%\end{figure}
	
	Of course, this interpretation implies that for $H_{c1}<H<H_{c2}$
	the system does {\em not} achieve full polarization. Exhibit 4
	(inset of panels (d) in Figs.~\ref{FIG:panelsA} and~\ref{FIG:panelsC}) puts
	this supposition to the test. These are direct Faraday force
	measurements of magnetization at $T=100$~mK from a $0.8$~mg sample. A small diamagnetic linear contribution was inferred from measurements above 8~T and subtracted from the data shown. For
	both geometries, the data show a divergent magnetic susceptibility
	at $H_{c1}$, seemingly followed by a complete saturation above. The
	blow-up of the high-field region shown in the insets in Figs.~\ref{FIG:panelsA} and~\ref{FIG:panelsC} reveal a different
	story. In order to reduce noise, the data for $\mu_{0}H>4$~T were binned
	together in $0.25$~T intervals. Between $\mu_{0}H_{c1}$
	and at least $5.5$~T the magnetization visibly lies \emph{below} the average
	value for all data above $8$~T (dashed line). The effect is small but
	clearly outside of the measurement uncertainty represented by the
	error bars.
	
	Incomplete polarization at high field could in principle stem from the small magnetic anisotropy present in the system~\cite{HagemansCaux_PRB_2005_XYZchainNosaturation,Chernyshev_PRB_2005_K2V3O8nosaturation}. However, this scenario can be excluded. The effect would have to be absent in the axial geometry, and in all cases confined to a much narrower field range, comparable to the spin flop field value~\cite{Chernyshev_PRB_2005_K2V3O8nosaturation}.
	The high field phase is indeed \emph{not} fully
	saturated. At the same time its polarization is in all
	cases over 98\%. Can such a highly polarized state support nematic
	order? Arguably it can, since the latter involves \emph{transverse}
	spin
	components~\cite{ShannonMomoi_PRL_2006_J1J2squarecircle,Smerald_PRB_2015_INSnematic}.
	As a reference, consider a classical AF at 99\% saturation. Despite
	being almost fully polarized it will still have an angle of as much
	as $16^\circ$ between consecutive spins, sufficient for AF order of
	transverse spin components. Similarly, such proximity to saturation
	leaves ample room for nematic correlations in  the transverse
	channel.
	
	There are stark discrepancies between our
	results and the powder measurements
	of Ref.~\cite{Skoulatos_PRB_2019_PutativeNematic}. In that
	work the magnetization curves were obtained by integrating AC
	susceptibility data. The one corresponding to the lowest
	experimental temperature $T=200$~mK is considerably less inflected
	below $H_{c1}$ than our curves shown in Figs.~\ref{FIG:panelsA}(d) and~\ref{FIG:panelsC}(d) or even those previously measured at $T=550$~mK~\cite{PovarovBhartiya_PRB_2019_BaCdVthermodynamics}.
	The inflection point that the authors take for the analogue of $H_{c1}$ doesn't
	correspond to any features in our  data for
	any field geometry. The reported value $\mu_{0}H_{c1}=3.78$~T is considerably lower
	than $\mu_{0}H_{c1}\gtrsim 4.0$~T observed in single crystals for \emph{all} geometries.
	The huge reduction of
	magnetization in the high-field phase concluded in Ref. ~\cite{Skoulatos_PRB_2019_PutativeNematic} is most certainly
	due to $H_{c1}$ being underestimated.

	The upper boundary assigned to the
	nematic phase in Ref.~\cite{Skoulatos_PRB_2019_PutativeNematic} does
	not correspond to any thermodynamic anomalies in single crystals.
	Counter-intuitively it moves out to higher fields with
	increased temperature. As is quite obvious from previously published single crystal data ~\cite{PovarovBhartiya_PRB_2019_BaCdVthermodynamics}, this simply indicates a broadening of the cusp in magnetization due to finite temperature. Our present data instead suggest that the upper phase boundary is more or
	less parallel to the lower one, as one would expect.  We
	cannot speculate on the origin of these discrepancies, except to
	note the intrinsic limitations of powder experiments in applied
	field and the difficulties of dealing with a dissipative component
	in AC susceptibility admitted by the authors of Ref.~\cite{Skoulatos_PRB_2019_PutativeNematic}.
	
	In summary, \BaCdV\ is {\it not} a square lattice model material as it was advertised to be. Instead, it has significantly alternating interactions along the $b$ direction. Nevertheless, it features strong FM-AFM frustration, the main ingredient for a pre-saturation bond-nematic phase. While the corresponding order parameter is fundamentally inaccessible to direct measurements, at the lowest temperatures, in a wide field range
	between $\mu_{0}H_{c1}\simeq 4.0$~T and $\mu_{0}H_{c2}\simeq 6.5$~T we find a
	novel \textit{almost} fully polarized thermodynamic phase. It does not support conventional (dipolar) magnetic order but  is fully consistent with our expectations for a quantum bond-nematic.

	This work was supported by the Swiss National Science Foundation,
	Division II. The instrument beamtime at D23 and IN12 was supported by the Swiss State Secretariat for Education, Research and Innovation (SERI) through a CRG Grant. We would like to acknowledge Dr. M. Reehius at HZB for his help with checking single crystals quality as well as the sample environment team of the ILL facility for their help with our experiments.
	
	\bibliography{The_Library}
	
\end{document}

\title{Supplemental material for ``Presaturation  phase with no dipolar order in a  quantum ferro-antiferromagnet''}

\author{V.~K.~Bhartiya}
   \email{vvivek@phys.ethz.ch}
    \affiliation{Laboratory for Solid State Physics, ETH Z\"{u}rich, 8093 Z\"{u}rich, Switzerland}

\author{K.~Yu.~Povarov}
        \affiliation{Laboratory for Solid State Physics, ETH Z\"{u}rich, 8093 Z\"{u}rich, Switzerland}

\author{D.~Blosser}
    \affiliation{Laboratory for Solid State Physics, ETH Z\"{u}rich, 8093 Z\"{u}rich, Switzerland}

\author{S.~Bettler}
    \affiliation{Laboratory for Solid State Physics, ETH Z\"{u}rich, 8093 Z\"{u}rich, Switzerland}

\author{Z.~Yan}
    \affiliation{Laboratory for Solid State Physics, ETH Z\"{u}rich, 8093 Z\"{u}rich, Switzerland}

\author{S.~Gvasaliya}
    \affiliation{Laboratory for Solid State Physics, ETH Z\"{u}rich, 8093 Z\"{u}rich, Switzerland}

\author{S.~Raymond}
   \affiliation{Univ. Grenoble Alpes, CEA, IRIG/DEPHY/MEM-MDN, F-38000 Grenoble, France}

\author{E.~Ressouche}
   \affiliation{Univ. Grenoble Alpes, CEA, IRIG/DEPHY/MEM-MDN, F-38000 Grenoble, France}

\author{K.~Beauvois}
   \affiliation{Institut Laue Langevin, 38000 Grenoble, France}

\author{J.~Xu}
   \affiliation{Helmholtz-Zentrum Berlin f\"{u}r Materialien und Energie GmbH, Hahn-Meitner-Platz 1, D-14109 Berlin, Germany}

\author{F.~Yokaichiya}
    \affiliation{Helmholtz-Zentrum Berlin f\"{u}r Materialien und Energie GmbH, Hahn-Meitner-Platz 1, D-14109 Berlin, Germany}

\author{A.~Zheludev}
    \email{zhelud@ethz.ch}
    \homepage{http://www.neutron.ethz.ch/}
    \affiliation{Laboratory for Solid State Physics, ETH Z\"{u}rich, 8093 Z\"{u}rich, Switzerland}

\date{\today}
\maketitle

\section{Structural phase transition}
\begin{figure}
	\centering
	\includegraphics[width=0.45\textwidth]{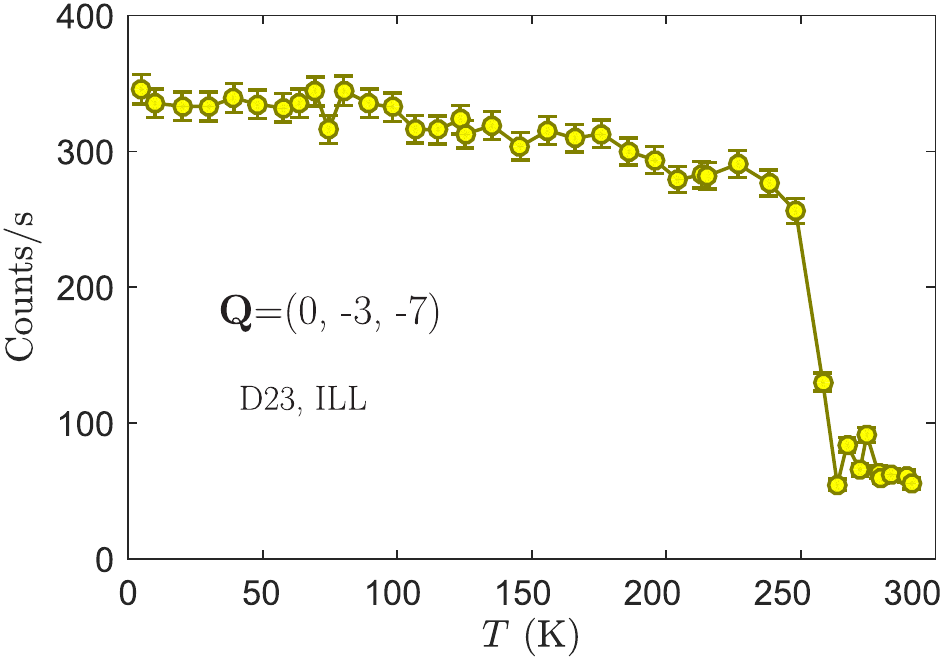}
	\caption{Temperature dependence of the  $(0,-3,-7)$ nuclear Bragg peak intensity obtained at  D23 neutron diffractometer (ILL) upon warming. The $(0,-3,-7)$ peak is extinct in $P_{bca}$ structure. }\label{FIG:ForPeak}
\end{figure}

As mentioned in the main text, a smoking gun evidence for the lowered symmetry of \BaCdV at low temperatures is the emergence of nominally extinct nuclear Bragg peaks around $250$~K. An example of the temperature dependence of the $(0,-3,-7)$ nuclear peak measured via neutron diffraction is shown in Fig.~\ref{FIG:ForPeak}. This data was measured on the CRG-D23 diffractometer at ILL on a $42$~mg sample installed in a $^4$He Orange cryostat. Neutrons with $k=2.6$~\AA$^{-1}$ were used. The observed temperature dependence is suggestive of crystal symmetry reduction.

We have solved the crystallographic structure of \BaCdV at $T=120$~K and indeed found it to be lower than $P_{bca}$. The low temperature structure corresponds to orthorhombic $P_{ca}2_{1}$ No. 29, a descendant of the original orthorhombic $P_{bca}$ No. 61.
The data ($14519$ reflections total) were obtained on a Bruker SMART-II single crystal X-ray diffractometer using a Mo-source and treated with the Brucker \textsc{AXS} software. Structure refinements were performed with \textsc{ShelX}. The results are presented in Table~\ref{TAB:LowT}. The listed values correspond to $R$-factors $R=0.054$, and $wR=0.16$.

One should also note that in $P_{ca}2_{1}$ notation (``dashed'' axes given in Table~\ref{TAB:LowT}) the $a$ and $c$ axis are swapped with respect to $P_{bca}$ notation. Nonetheless, for consistency with the previous work on \BaCdV we keep referring to $c$ as the longest and to $a$ as the shortest axes. These are the axes non-dashed labels in Table~\ref{TAB:LowT}.

\begin{figure}
	\centering
	\includegraphics[width=0.5\textwidth]{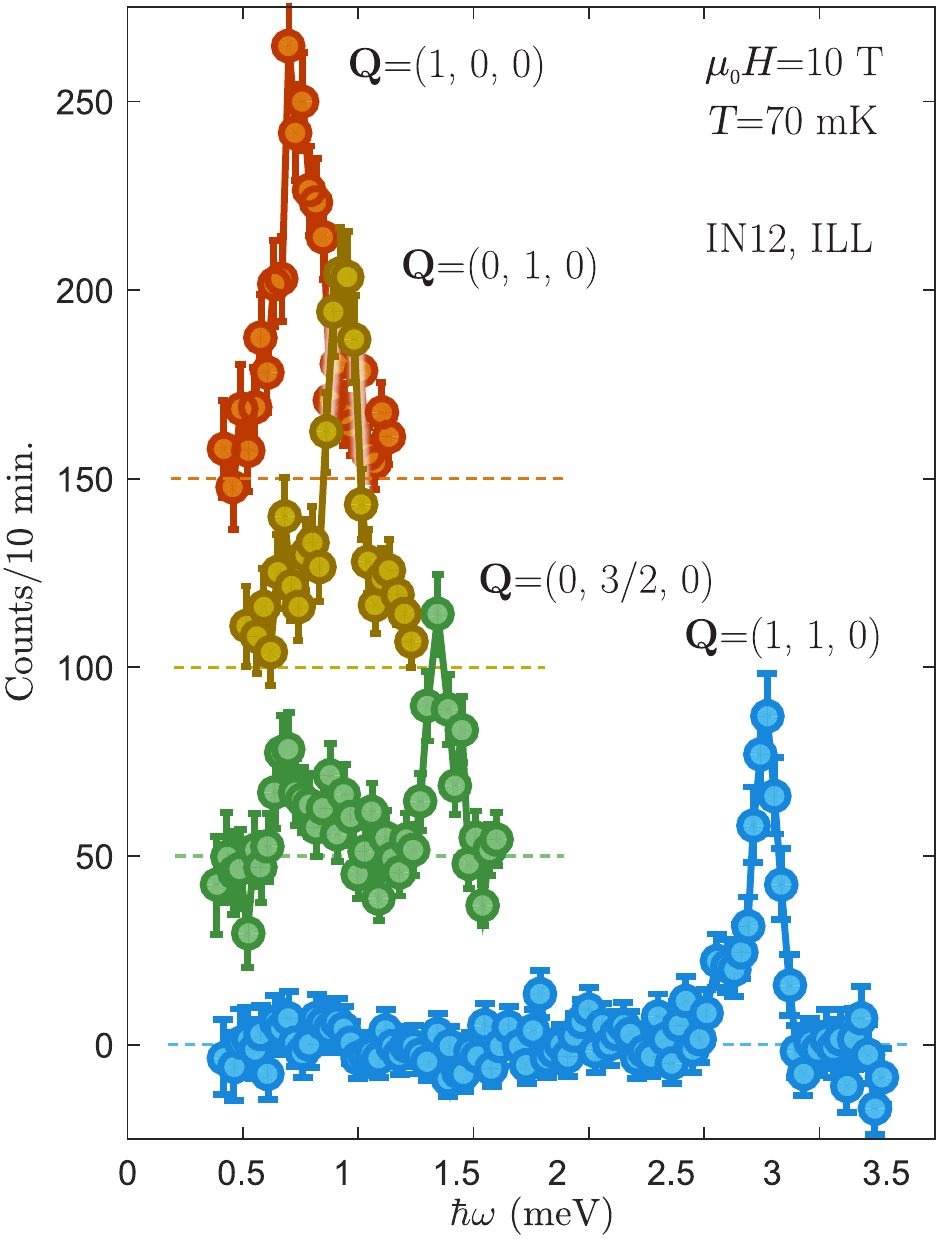}
	\caption{A few representative, background subtracted, inelastic neutron scattering constant-$\mathbf{Q}$ scans in the fully polarized phase ($10$~T $\mathbf{H}\parallel\mathbf{c}$) of \BaCdV at $70$~mK.  An offset of $50$~counts/curve is introduced for better visibility and their respective zeroes are indicated by dashed lines.}\label{FIG:PrettyScans}
\end{figure}
\begin{table}
  \centering
  \begin{tabular}{ l c c c c }
  \hline\hline
  Lattice parameters & $a'$ ($c$) & $b'$ ($b$) & $c'$ ($a$)\\
  \hline
  &  18.8591(9) & 8.8911(4) & 8.8621(4)\\
  & & & \\
    \hline
        Symmetry transformations & & & $P_{ca}2_{1}$ \\
    \hline
        & $x$ & $y$ & $z$ \\
    & $-x$ & $-y$ & $\frac{1}{2}+z$ \\
    & $\frac{1}{2}+x$ & $-y$ & $z$ \\
    & $\frac{1}{2}-x$ & $y$ & $\frac{1}{2}+z$ \\
    & & & \\
    \hline
      Atom & $x$ & $y$ & $z$ & U$_{\text{iso}}$\\
      \hline
    Ba 1 & 0.58655(2) & 0.73248(3)  & -0.20183(4)  & 0.00460(4) \\
    Ba 2 & 0.91837(2) & 0.24762(3)  &  0.43979(4)  & 0.00465(4) \\
    Cd 1 & 0.59795(2) & 0.60137(4)  &  0.27429(5)  & 0.00457(5) \\
    Cd 2 & 0.41362(2) & 0.86375(4)  & -0.05557(5)  & 0.00470(5) \\
    V 1  & 0.73144(5) & 0.39048(9)  &  0.69354(10) & 0.00343(9) \\
    V 2  & 0.78961(5) &-0.11011(9)  &  0.69465(10) & 0.00317(9) \\
    P 1  & 0.46975(8) & 0.52771(16) &  0.04790(18) & 0.00401(17) \\
    P 2  & 0.75096(8) & 0.63556(14) &  0.44851(19) & 0.00289(15) \\
    P 3  & 0.46817(8) & 1.04426(16) & -0.35171(17) & 0.00347(17) \\
    P 4  & 0.74626(8) & 0.13737(14) &  0.44741(19) & 0.00331(16) \\
    O 1  & 0.4580(2)  & 0.8816(4)   & -0.2934(5)   & 0.0043(4) \\
    O 2  & 0.7978(2)  & 0.5222(4)   &  0.3610(5)   & 0.0050(5) \\
    O 3  & 0.7040(3)  &-0.1071(5)   &  0.7010(6)   & 0.0080(6) \\
    O 4  & 0.8029(2)  & 0.7475(4)   &  0.5191(5)   & 0.0053(5) \\
    O 5  & 0.4614(2)  & 1.0429(5)   & -0.5252(5)   & 0.0063(5) \\
    O 6  & 0.5518(2)  & 0.5426(5)   &  0.0568(5)   & 0.0066(5) \\
    O 7  & 0.8014(2)  & 0.0598(4)   &  0.5474(5)   & 0.0048(4) \\
    O 8  & 0.4352(3)  & 0.5629(5)   &  0.1993(6)   & 0.0083(6) \\
    O 9  & 0.4528(2)  & 0.3614(5)   &  0.0026(5)   & 0.0061(5) \\
    O 10 & 0.7036(2)  & 0.5575(5)   &  0.5636(5)   & 0.0054(5) \\
    O 11 & 0.6922(2)  & 0.0299(4)   &  0.3719(5)   & 0.0051(5) \\
    O 12 & 0.6994(2)  & 0.2489(4)   &  0.5340(5)   & 0.0057(5) \\
    O 13 & 0.4449(2)  & 0.6250(5)   & -0.0881(5)   & 0.0061(5) \\
    O 14 & 0.7894(2)  & 0.2215(4)   &  0.3268(5)   & 0.0050(5) \\
    O 15 & 0.5439(3)  & 1.0986(5)   & -0.3202(6)   & 0.0085(6) \\
    O 16 & 0.6995(2)  & 0.7138(4)   &  0.3373(5)   & 0.0046(4) \\
    O 17 & 0.8167(3)  & 0.3963(5)   &  0.6871(6)   & 0.0078(5) \\
    O 18 & 0.4097(2)  & 1.1432(5)   & -0.2802(6)   & 0.0070(5) \\
   \hline\hline
  \end{tabular}
  \caption{Chemical structure of the low-temperature phase of \BaCdV. Correspondence between the axis labels for low- and high temperature structures is shown in brackets.}\label{TAB:LowT}.
\end{table}

\section{Inelastic neutron scattering}
The inelastic neutron scattering measurements were performed on the CRG-IN12 cold neutron 3-axis spectrometer at ILL. The 320~mg sample consisted of two \BaCdV crystals with 98.8\% isotopically enriched 144Cd, co-aligned with a total mosaic of about $10^\circ$ in the $(h,k,0)$ scattering  plane on an aluminium sample holder. The assembly was mounted on the cold finger of the $^3$He-$^4$He dilution refrigerator and installed on IN12 with a vertical cryomagnet. The measurements were performed at the base temperature of $70$~mK and $10$~T magnetic field $\mathbf{H}\parallel\mathbf{c}$. Examples of constant-$\mathbf{Q}$ inelastic scans contributing to the dispersion in Figure~1 of the main text are shown in Fig.~\ref{FIG:PrettyScans}. The data collected on IN12 are available at https://doi.ill.fr/10.5291/ILL-DATA.CRG-2577.